\documentclass[11pt]{article}
\usepackage{color,latexsym,amsmath,amssymb,path,enumerate,url,fullpage}

\newcommand{\ignore}[1]{}

\newtheorem{theorem}{Theorem}[section]
\newtheorem{lemma}[theorem]{Lemma}
\newtheorem{corollary}[theorem]{Corollary}

\newcommand{\Proof}[1]
        {
        \noindent
        \emph{Proof #1.}~
        }
\newsavebox{\smallProofsym}                     
\savebox{\smallProofsym}                        %
        {
        \begin{picture}(7,7)                    %
        \put(0,0){\framebox(6,6){}}             %
        \put(0,2){\framebox(4,4){}}             %
        \end{picture}                           %
        }                                       %
\newcommand{\smalleop}[1]
        {
        \mbox{} \hfill #1~~\usebox{\smallProofsym}\!\!\!\!\!\!\
        }
\newenvironment{theProof}[1]
        {
        \Proof{#1}}{\smalleop{}
        \medskip

        }
\usepackage{graphicx,psfrag}
\usepackage[rflt]{floatflt}

\newcommand{\placefig}[2]
        {\includegraphics[width=#2]{#1.eps}}
\usepackage{dsfont}

\newcommand{\tri}[1]{{\mathsf{tr}}(#1)}

\newcommand{\tr}[1]{{\cal T}(#1)}
\newcommand{\pmt}[1]{{\cal M}(#1)}

\newcommand{\scy}[1]{{\cal C}(#1)}

\newcommand{\pmtc}[1]{{\mathsf{pm}}(#1)}

\newcommand{\scyc}[1]{{\mathsf{sc}}(#1)}
\newcommand{\scyct}[1]{{\mathsf{sc}_{_{\Delta}}}(#1)}

\newcommand{\supp}[1]{\mathsf{supp}(#1)}

\newcommand{\flip}[1]{\mathsf{flip}(#1)}
\newcommand{\flipt}[1]{\mathsf{flip}_T(#1)}

\begin{document}
\pagenumbering{arabic}
\date{}

\title{Counting Plane Graphs: Perfect Matchings, Spanning Cycles, and Kasteleyn's Technique\thanks{%
Work on this paper by the first two authors was partially supported by Grant 338/09 from
the Israel Science Fund. Work by Micha Sharir was also
supported by NSF Grant CCF-08-30272,
by Grant 2006/194 from the U.S.-Israel Binational Science Foundation,
and by the Hermann
Minkowski--MINERVA Center for Geometry at Tel Aviv University.
Emo Welzl acknowledges support from the EuroCores/EuroGiga/ComPoSe SNF grant 20GG21\_134318/1.
Part of the work on this paper was done at the Centre Interfacultaire Bernoulli (CIB),
during the Special Semester on Discrete and Computational Geometry, Fall 2010, and supported by the Swiss National Science Foundation}}

\author{
Micha Sharir\thanks{%
School of Computer Science, Tel Aviv University,
Tel Aviv 69978, Israel and Courant Institute of Mathematical
Sciences, New York University, New York, NY 10012, USA.
{\sl michas@tau.ac.il} }
\and
Adam Sheffer\thanks{%
School of Computer Science, Tel Aviv University,
Tel Aviv 69978, Israel.
{\sl sheffera@tau.ac.il}
}
\and
Emo Welzl \thanks{Institute for Theoretical Computer Science,
ETH Z\"urich, CH-8092 Z\"urich, Switzerland.
{\sl welzl@inf.ethz.ch}
}}

\maketitle

\begin{abstract}
We derive improved upper bounds on the number of crossing-free straight-edge spanning cycles (also known as Hamiltonian tours and simple polygonizations) that can be embedded over any specific set of $N$ points in the plane.
More specifically, we bound the ratio between the number of spanning cycles (or perfect matchings) that can be embedded over a point set and the number of triangulations that
can be embedded over it. The respective bounds are $O(1.8181^N)$ for cycles and $O(1.1067^N)$ for matchings.
These imply a new upper bound of $O(54.543^N)$ on the number of crossing-free straight-edge spanning cycles that can be embedded over any specific set of $N$ points in the plane
(improving upon the previous best upper bound $O(68.664^N)$).
Our analysis is based on Kasteleyn's linear algebra technique.
\end{abstract}

\section{Introduction}

Finding a Hamiltonian cycle in a graph is $\mathbf{NP}$-complete even if the graph is known to be planar \cite{GJT76}.
\emph{Counting} the number of Hamiltonian cycles that are contained in a graph is $\mathbf{\#P}$-complete even if the graph is known to be planar and all of its vertices have
degree three \cite{LOT03}. In this paper we consider a more general case (and thus, probably more difficult) concerning the number of all crossing-free straight-edge Hamiltonian cycles
that can be embedded over a specific set of points in the plane. That is, given a set $S$ of labeled points in the plane, we consider the number of Hamiltonian cycles that have a straight-edge planar embedding over $S$.

We only consider the problem of obtaining an upper bound for the number crossing-free straight-edge Hamiltonian cycles that can be embedded over a set
of $N$ points in the plane.
To do this, we rely on Kasteleyn's technique \cite{Kast67}, and on edge-flipping techniques that were developed in a previous paper by the authors \cite{HSSTW11}.
No familiarity with \cite{HSSTW11} is necessary, since we re-introduce all the notions that we require from it.
We now give a detailed and more formal definition of the problem.

A {\em planar graph} is a graph that can be embedded in the plane in such
a way that its vertices are embedded as points and its edges are embedded as Jordan arcs that connect the respective pairs of points and can meet only at a common endpoint.
A {\em crossing-free straight-edge graph} is a plane embedding of a planar graph
such that its edges are embedded as non-crossing straight line
segments. In this paper, we only consider crossing-free straight-edge graphs.
Moreover, we only consider embeddings where the points are in general position, that is,
where no three points are collinear. (For upper bounds on the number of graphs, this involves no loss of generality, because the number of graphs can only grow
when a degenerate point set is slightly perturbed into general position.)
For simplicity, we sometimes refer to such graphs as \emph{plane graphs}.

We focus on upper bounding the maximal number of plane \emph{spanning cycles} (also known as \emph{Hamiltonian cycles},
\emph{Hamiltonian tours}, and \emph{simple polygonizations}) that can be embedded over a fixed set of points in the plane.
For a set $S$ of points in the plane, we denote by $\scy{S}$ the set of all crossing-free straight-edge spanning cycles of $S$,
and put $\scyc{S}:= \left|\scy{S}\right|$.
Moreover, we let $\scyc{N}=\max_{|S|=N}\scyc{S}$. So, in other words, the main goal of this paper is to obtain sharp upper bounds on $\scyc{N}$.

There are many similar variants of this problem, such as bounding the number of plane forests, spanning trees, triangulations, and general plane graphs.
Recent work on some of these variants can be found in \cite{AHHHKV07,HSSTW11,SS10}, and we try to keep a comprehensive list of the up-to-date bounds in a dedicated webpage\footnote{\url{http://www.cs.tau.ac.il/~sheffera/counting/PlaneGraphs.html} (version of June 2010).}.
It seems that the case of spanning cycles is the most popular one, already considered in \cite{ACNS82,A79,BKKSS07,DSST11,GNT00,NM80,SW06} and many others.
Moreover, spanning cycles were the first case for which bounds were published, namely the bounds $3/20 \cdot 10^{N/3} \le \scyc{N} \le 2\cdot6^{N-2} \cdot (\lfloor N/2 \rfloor)!$ in \cite{NM80}.
A brief history of the steady progress on bounding the number of spanning cycles can be found in a dedicated webpage by Erik Demaine\footnote{\url{http://erikdemaine.org/polygonization/} (version of June 2010).}.
Currently, the best known lower bound is $\scyc{N} = \Omega(4.642^N)$, due to Garc\'ia, Noy, and Tejel \cite{GNT00},
and the previous upper bound is $\scyc{N} = O(68.664^N)$ by Dumitrescu~{\em et al.}~\cite{DSST11}.
We derive the improved bound $\scyc{N}  = O(54.543^N)$.

These problems have also been studied from an algorithmic point of view, deriving algorithms for enumeration or counting of the plane graphs (or other graph types) that can be embedded over a given point set (such as in \cite{KT08,RW09}).
The combinatorial upper bounds are useful for analyzing the running times of such algorithms, and also to answer questions such as ``how many bits are required to represent a triangulation
(or any other kind of plane graphs)?".

Our bound (as do some of the previous bounds) relies on \emph{triangulations}.
A triangulation of a set $S$ of $N$ points in the plane is a maximal plane graph on $S$
(that is, no additional straight edges can be inserted without crossing some of the existing edges).
For a set $S$ of points in the plane, we denote by $\tr{S}$ the set of all
triangulations of $S$, and put $\tri{S}:= \left|\tr{S}\right|$.
Moreover, we let $\tri{N}=\max_{|S|=N}\tri{S}$.
Currently, the best known bounds for $\tri{N}$ are $\tri{N}<30^N$ \cite{SS10},
and $\tri{N} = \Omega (8.65^N)$ \cite{DSST11}.

The upper bound by Dumitrescu~{\em et al.}~\cite{DSST11} is obtained by proving that for every set $S$ of $N$ points in the plane $\scyc{S} = O\left(2.2888^N \right) \cdot \tri{S}$.
This has sharpened an earlier bound of Buchin~{\em et al.}~\cite{BKKSS07}, who showed that every triangulation $T$ of $S$ contains at most $30^{N/4} \approx 2.3404^N$ spanning cycles
(i.e., cycles whose edges belong to $T$), implying\footnote{The implication comes from the fact that every spanning cycle, and in fact every plane graph, is contained in at least one triangulation; see Section \ref{sec:pre}.}
that $\scyc{S} < 2.3404^N \cdot \tri{S}$.
Combining the above ratio with the bound $\tri{N}<30^N$ directly implies the asserted bound. We derive our bound in a similar manner, showing that $\scyc{S} = O\left(1.8181^N \right) \cdot \tri{S} = O(54.5430^N)$.

\begin{figure}[h]
\centerline{\placefig{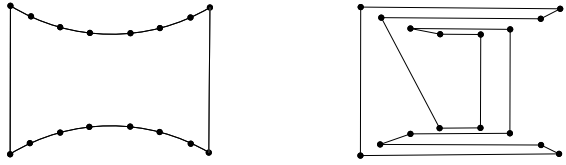}{0.45\textwidth}}
\vspace{-1mm}

\caption{\small \sf Two spanning cycles embedded over a double chain point configuration.}
\label{fi:doubleCpol}
\vspace{-2mm}
\end{figure}

In spite of our improved bound, we strongly believe, and conjecture, that for every point set $S$ (of size at least $N_0$, for some constant $N_0$) one has $\scyc{S} < \tri{S}$, and perhaps even a much sharper ratio holds.
The best lower bound for this ratio that we know of is obtained from the \emph{double chain configuration}, presented in \cite{GNT00}
(and depicted in Figure \ref{fi:doubleCpol}). It is shown in \cite{GNT00} that when $S$ is a double chain configuration, $\tri{S}=\Theta^*\left(8^N\right)$ and
$\scyc{S}=\Omega^*\left(4.64^N\right)$.\footnote{In the notations $O^*()$, $\Theta^*()$, and $\Omega^*()$, we neglect polynomial factors.}
Thus, in this case, $\scyc{S} / \tri{S} = \Omega^*(0.58^N)$. (It is \emph{stated} in \cite{AHHHKV07}, albeit without proof, that $\scyc{S}=O(5.61^N)$, so this example supports our conjecture.)

In Section \ref{sec:pre} we go over the preliminaries required for our analysis. These include, among others, the edge-flip techniques used in \cite{HSSTW11}.
Section \ref{se:firstB} derives the bound $\scyc{S} = O\left(12^{N/4}\right) \cdot \tri{S} = O\left(1.8613^N\right) \cdot \tri{S}$ for any set $S$ of $N$ points in the plane.
As part of this derivation, we describe Kasteleyn's technique for counting perfect matchings and present a new way of applying it.
In Section \ref{sec:matchings} we use the same methods to prove an upper bound on the ratio between the number of plane perfect matchings and the number of triangulations,
showing that $\pmtc{S}  = O(1.1067^N) \cdot \tri{S}$ (where $\pmtc{S}$ is the number of matchings that can be embedded over the point set $S$).
Finally, Section \ref{sec:improved} contains a more complex analysis of spanning cycles,
implying the improved bound $\scyc{S} = O\left(10.9247^{N/4}\right) \cdot \tri{S} = O\left(1.8181^N\right) \cdot \tri{S}$.

\section{Preliminaries} \label{sec:pre}
In this section we establish some notations and lemmas that are required for the following sections.
\begin{list}{}{\leftmargin= 0em}
\item Given two plane graphs $G$ and $H$ over the same point set $S$, if every edge of $G$ is also an edge of $H$, we write $G \subseteq H$.
\item Hull edges and vertices (resp., interior edges and vertices) of a graph embedded on a point set $S$ are those that are part of the boundary of the convex hull of $S$ (resp.,
not part of the convex hull boundary).
\item Given a set $S$ of $N$ points in the plane, we denote by $h$ the number of hull vertices of $S$, and put $n=N-h$, which is the number of vertices in $S$ interior
to its convex hull.
\end{list}

\subsection{The support of a graph}
Let us denote by $\scyct{N}$ the maximal number of plane spanning cycles that can be contained in any fixed triangulation of a set of $N$ points in the plane.
Moreover, denote the set of spanning cycles contained in a triangulation $T$ by $\scy{T}$, so $\scyct{N}=\max_{|S|=N, \ T \in \tr{S}}|\scy{T}|$.

Any spanning cycle (or, for that matter, any plane graph) is contained in at least one triangulation. Therefore, we can upper bound the number of spanning cycles
of a set $S$ of $N$ points in the plane by going over every triangulation $T \in \tr{S}$ and counting the number of spanning cycles contained in $T$.
This implies the bound $\scyc{N} \le \tri{N} \cdot \scyct{N}$.
Applying the bounds $\tri{N} < 30^N$ from \cite{SS10} and $\scyct{N} \le 30^{N/4}$ from \cite{BKKSS07}, we obtain $\scyc{N} < 30^{5N/4} \approx 70.21^N$.

This bounding method seems rather weak since it potentially counts some spanning cycles many times.
For example, consider a spanning cycle consisting of two convex chains facing each other, as depicted in the left-hand side of Figure \ref{fi:doubleCpol}.
Garc\'ia, Noy, and Tejel \cite{GNT00} show that such a spanning cycle is contained in $\Theta^*(8^N)$ triangulations of its point set.
Therefore, the above method will count this spanning cycle $\Theta^*(8^N)$ times. However, as stated in \cite{AHHHKV07}, this point set has only $O(5.61^N)$ spanning cycles.

In order to deal with this inefficiency, we define the notion of \emph{support} (the same notion was also used in \cite{DSST11,HSSTW11,SS10,SSW10,SW06b}).
Given a plane edge graph $G$ embedded over a set $S$ of points in the plane, we say that $G$ has a support of $x$ if $G$ is contained in
(exactly) $x$ triangulations of $S$; we write $\supp{G}=x$. Notice that
\begin{equation} \label{eq:supp}
\scyc{S} = \sum_{T \in \tr{S}} \sum_{C \in \scy{T}}\frac{1}{\supp{C}},
\end{equation}
because every spanning cycle $C$ contributes exactly one to the right side of the equation
(it appears in $\supp{C}$ terms of the first sum, and contributes $1/\supp{C}$ in every appearance). We will use (\ref{eq:supp}) to obtain
better upper bounds for $\scyc{N}$, by showing that, on average, $\supp{C}$ is large.

\subsection{Ps-flippable edges}

\begin{figure}[h]
\psfrag{aa}[cc]{(a)}
\psfrag{bb}[cc]{(b)}
\psfrag{cc}[cc]{(c)}
\psfrag{a}{$a$}
\psfrag{b}{$b$}
\psfrag{c}{$c$}
\psfrag{d}{$d$}
\psfrag{e}{$e$}
\psfrag{a2}{$a$}
\psfrag{b2}{$b$}
\psfrag{c2}{$c$}
\psfrag{d2}{$d$}
\psfrag{e2}{$e$}

\centerline{\placefig{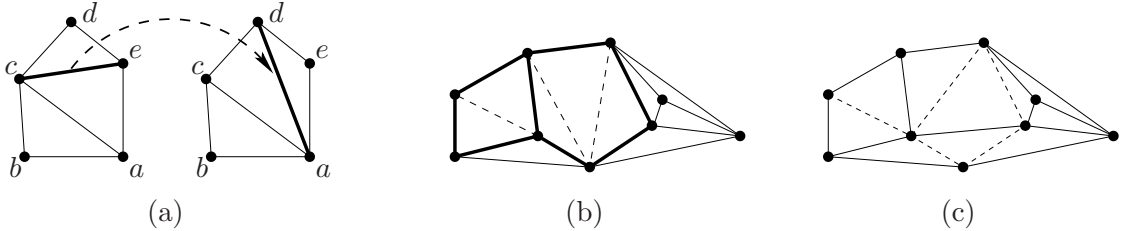}{0.9\textwidth}}
\vspace{-1mm}

\caption{\small \sf (a) The edge $ce$ can be flipped to the edge $ad$.
(b) A set of three (dashed) ps-flippable edges which are diagonals of interior-disjoint convex quadrilateral and convex pentagon.
(c) A convex decomposition obtained by removing the dashed edges from the triangulation; here too the dashed edges are diagonals of
pairwise disjoint convex faces, and form a set of ps-flippable edges.}
\label{fi:flip}
\vspace{-2mm}
\end{figure}

An edge in a triangulation is said to be {\em flippable}, if its two incident triangles form a convex quadrilateral.
A flippable edge can be {\em flipped}, that is, removed from the
graph of the triangulation and replaced by the other diagonal of the corresponding quadrilateral.
Such an operation is depicted in Figure \ref{fi:flip}(a), where the edge $ce$ can be flipped to the edge $ad$.

In \cite{HSSTW11}, we present the concept of \emph{pseudo simultaneously flippable edges} (or \emph{ps-flippable edges}, for short).
Given a triangulation $T$, we say that a subset $F$ of its edges is a set of ps-flippable edges if the edges of $F$
are diagonals of interior-disjoint convex polygons (whose boundaries are also parts of $T$). For example, in Figure \ref{fi:flip}(b), the three dashed edges form a set of ps-flippable edges,
since they are diagonals of interior-disjoint convex quadrilateral and convex pentagon (another set of ps-flippable edges, in a different triangulation, is depicted in Figure \ref{fi:flip}(c)).

Ps-flippable edges are related to \emph{convex decompositions}. A convex decomposition of a point set $S$ is a crossing-free straight-edge graph $D$ on $S$
such that (i) $D$ includes all the hull edges, (ii) each bounded face of $D$ is a convex polygon, and (iii) no point of $S$ is isolated in $D$.
See Figure \ref{fi:flip}(c) for an illustration. For additional information about convex decompositions, see, for example, \cite{H09}.
Notice that if $T$ is a triangulation that contains $D$, the edges of $T \setminus D$ form a set of ps-flippable edges, since they are the
diagonals of the interior-disjoint convex polygons of $D$ (again, consider the dashed edges in Figure \ref{fi:flip}(c) for an illustration).
Thus, finding a large set of ps-flippable edges in a triangulation $T$ is equivalent to
finding a convex decomposition with a small number of faces (or edges) in $T$.

In \cite{HSSTW11}, we prove the two following lemmas.
\begin{lemma} \label{le:psFlip}
Every triangulation $T$ over a set of $N$ points in the plane contains a set $F$ of $N/2-2$ ps-flippbale edges.
Also, there are triangulations with no larger sets of ps-flippable edges.
\end{lemma}

\begin{lemma} \label{le:psSupport}
Consider a triangulation $T$, a set $F$ of $N/2-2$ ps-flippable edges in $T$, and a graph $G \subseteq T$.
If $G$ does not contain $j$ edges from $F$ then $\supp{G} \ge 2^j$.
\end{lemma}
\begin{theProof}{sketch}
Consider the  set $F'=F\setminus G$ of $j$ ps-flippable edges.
The convex faces of $T\setminus F'$ can be triangulated in at least $2^j$ ways (the bound is tight when every edge of $F'$ is a diagonal of a distinct quadrangular face of $T\setminus F'$),
and each of the resulting triangulations contains $G$. See \cite{HSSTW11} for more details.

\end{theProof}

\begin{figure}[h]
\centerline{\placefig{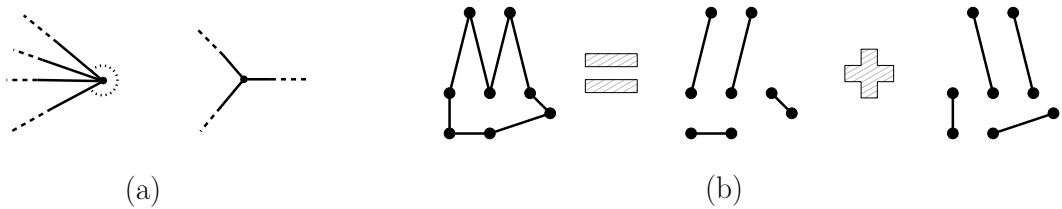}{0.85\textwidth}}
\vspace{-1mm}

\caption{\small \sf (a) A vertex is valid if and only if it is not a reflex vertex of any face.
(b) In a point set of an even size, every spanning cycle is the union of two edge-disjoint perfect matchings.}
\label{fi:valid}
\vspace{-2mm}
\end{figure}

We now describe another property of convex decompositions (not discussed in \cite{HSSTW11}).
Consider a set $S$ of points in the plane and a crossing-free straight-edge graph $G$ embedded on $S$. We say that an interior point $p\in S$
has a \emph{valid triple of edges in $G$} if there exist three points
$a,b,c \in S$ such that $p$ is contained in the convex hull of $\{a,b,c\}$ and the edges $ap$, $bp$, and $cp$ belong to $G$.
To simplify the notation, we refer to vertices with valid triples as \emph{valid} (with respect to $G$), and to the other interior vertices as \emph{non-valid}.
\begin{lemma} \label{le:validTri}
Let $S$ be a set of points in the plane and let $G$ be a crossing-free straight-edge graph over $S$ that contains all the edges of the convex hull of $S$.
Then $G$ is a convex decomposition of $S$ if and only if every interior vertex of $S$ is valid with respect to $G$.
\end{lemma}
\begin{theProof}{\!\!}
An interior vertex $v$ is a reflex vertex of some face of $G$ if and only if $v$ is non-valid
(an example is depicted in Figure \ref{fi:valid}(a)).
The lemma follows by observing that $G$ is a convex decomposition if and only if no bounded face of $G$ has a reflex vertex.
\end{theProof}

\subsection{Spanning cycles and perfect matchings}
Our analysis, as most of the previous works dealing with the number spanning cycles, heavily relies on the number of plane
perfect matchings on $S$ (for example, see \cite{BKKSS07,DSST11,SW06}). To see the connection between the two problems,
notice that if $|S|$ is even, every spanning cycle $C$ is the union of two edge-disjoint perfect matchings on $S$;
namely, the matching consisting of the even-indexed edges of $C$, and the matching consisting of the odd-indexed edges.
An illustration of this property is depicted in Figure \ref{fi:valid}(b).
Denote by $\pmt{S}$ the set of all plane perfect matchings on $S$, and put $\pmtc{S}=|\pmt{S}|$.
We also set $\pmtc{N}=\max_{|S|=N}\pmtc{S}$.
Hence, a simple upper bound on $\scyc{S}$ is $\pmtc{S}^2$.
In general, the union of two edge-disjoint perfect matchings is not always a spanning cycle,
but it is a cover of $S$ by vertex-disjoint even-sized cycles.

To deal with point sets of odd size, we use the following lemma:
\begin{lemma} \label{le:even}
Let $c>1$ be a constant such that every set $S$ of an even number of points in the plane satisfies $\scyc{S} = O(c^{|S|})$.
Then $\scyc{S} = O(c^{|S|})$ also holds for sets $S$ of an odd number of points.
\end{lemma}
\begin{theProof}{\!\!}
Consider a set $S$ of $N$ points in the plane, where $N$ is odd.
Pick a new point $p$ outside the convex hull of $S$, and put $S'=S \cup \{p\}$. Let $C$ be a plane spanning cycle of $S$.
Then there exists an edge $e=vu$ of $C$ such that $p$ can be connected to the two endpoints $u,v$ of $e$ without crossing $C$.
Indeed, this is a projective variant of the property, noted in \cite{GY80},
that every finite collection of non-crossing straight segments in the plane contains a segment $e$
such that no other segment lies vertically above any point of $e$ (see also \cite[Section 8.7]{OR98}).
By replacing $e$ with the edges $vp$ and $pu$, we obtain a crossing-free spanning cycle of $S'$.
This implies that we can map every spanning cycle of $S$ to
a distinct spanning cycle of $S'$, and thus, $\scyc{S} \le \scyc{S'}$.
The lemma then follows since $\scyc{S'} = O(c^{N+1}) = O(c^N)$. \end{theProof}

Bounding the number of perfect matchings on $S$ within a fixed triangulation $T$ can be done by the beautiful linear-algebra technique of Kasteleyn \cite{Kast67},
described in detail in \cite[Section 8.3]{LP86}; see Section \ref{se:firstB} for more details. Buchin~{\em et al.}~\cite{BKKSS07} have used this technique to show that any triangulation
$T$ of $S$ contains at most $6^{N/4}$ perfect matchings, and at most $30^{N/4} \approx 2.3403^N$ spanning cycles. We also note that Sharir and Welzl \cite{SW06} showed that
$\pmtc{S} = O\left(10.05^N\right)$, completely bypassing the approach of counting matchings (or other graphs) \emph{within} a triangulation.

\section{A first bound} \label{se:firstB}

In this section we first review an enhanced variant of Kasteleyn's technique and then use it to derive the following initial upper bound on $\scyc{S}$.

\begin{theorem}
\label{th:scVtr}
For any set $S$ of $N$ points in the plane,
\[ \scyc{S} = O\left(12^{N/4}\right) \cdot \tri{S} = O\left(1.8613^N\right) \cdot \tri{S}. \]
\end{theorem}
\begin{theProof}{\!\!}
First, by Lemma \ref{le:even}, we may assume that $N$ is even. Consider a triangulation $T$ of $S$. As already observed,
every spanning cycle contained in $T$ is the union of two edge-disjoint perfect matchings contained in $T$.
Given a plane graph $G$, we denote by $\pmt{G}$ the set of all perfect matchings that are contained in $G$.
Recalling (\ref{eq:supp}), we have \[ \scyc{S} \le \sum_{T \in \tr{S}} \sum_{M_1,M_2 \in \pmt{T} \atop M_1,M_2 \text{ edge-disjoint}} \frac{1}{\supp{M_1 \cup M_2}}. \]

\begin{figure}[h]
\centerline{\placefig{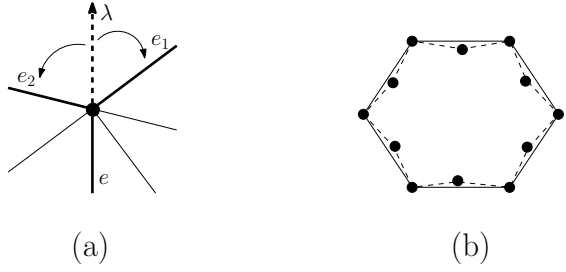}{0.45\textwidth}}
\vspace{-1mm}

\caption{\small \sf (a) The set $\{e,e_1,e_2\}$ is a valid triple of edges. (b) A set of 12 points in a double circle configuration.}
\label{fi:triple}
\vspace{-2mm}
\end{figure}

\noindent (The inequality comes from the fact that not every pair $M_1,M_2$ of matchings, as in the sum, necessarily yields a spanning cycle.)
Let us fix the ``first"  perfect matching $M_1 \subset T$; as mentioned above, Buchin~{\em et al.}~\cite{BKKSS07} prove that $|\pmt{T}| \le 6^{N/4}$,
so there are at most $6^{N/4}$ choices of $M_1$.
Next, we construct a convex decomposition $D$ such that $M_1 \subset D \subset T$, as follows.
We start with $M_1$ and add all the missing hull edges; let us denote the resulting graph as $D'$.
By Lemma \ref{le:validTri}, it suffices to ensure that every interior point $p\in S$ is connected in $D$ to (at least) three points $a,b,c \in S$, such that $p$ is inside the convex hull of $\{a,b,c\}$. Every interior vertex $p$ of $S$ has degree 1 in $D'$, so we start by setting $D:= D'$, and then, for each interior point $p \in S$,
we add to $D$ two additional edges of $T$ adjacent to $p$, so as to create a valid triple.
To do so, let $e$ be the edge of $D'$ (that is, of $M_1$) incident to $p$, and let $\lambda$ be the ray emanating from $p$ in the opposite direction.
Let $e_1$ (resp., $e_2$) be the first edge of $T$ incident to $p$ encountered in clockwise (resp., counterclockwise) direction from $\lambda$;
see Figure \ref{fi:triple}(a). Then $\{e,e_1,e_2\}$ is a valid triple of edges, and we add $e_1,e_2$ to $D$.
After applying this step to each interior point $p$, the resulting graph $D$ is indeed a convex decomposition of $S$.

We denote by $F$ the set of edges that are in $T$ but not in $D$. The edges of $F$ are diagonals of interior-disjoint
convex polygons, and thus $F$ is a set of ps-flippable edges.
By Euler's formula, the triangulation $T$ contains $3N-2h-3$ interior edges, and $D$ contains at most $2n+N/2$ interior edges (at most $N/2$ edges of $M_1$ and $2n$ added edges to form $n$ valid triples).
Therefore, \[ |F| \ge 3N-2h-3 - (2n+N/2) = N/2-3. \]
{\bf Remark.} Note the strength of this bound: Lemma \ref{le:psFlip} has a rather involved proof, given in \cite{HSSTW11},
and it yields a set of $N/2-2$ ps-flippable edges in the entire triangulation.
In contrast, here we get the same number (minus 1) after we remove from $T$ an arbitrary perfect matching, with a considerably simpler analysis.
Thus the significance of the analysis in \cite{HSSTW11} (involving Lemma \ref{le:psFlip}) is only for triangulations which contain no perfect matching on $S$. For example, any triangulation with
more than $N/2$ interior vertices of degree 3 cannot contain a perfect matching, since, as is easily checked, two interior vertices of degree 3 cannot share an edge. \vspace{1mm}

Without loss of generality, we assume that $F$ consists of \emph{exactly} $N/2-3$ edges. We now proceed to bound the number of ways to choose the second matching $M_2$ while taking the supports of the resulting graphs $M_1 \cup M_2$ into account.
Since $M_1$ and $M_2$ have to be edge-disjoint, we can remove the $N/2$ edges of $M_1$ from $T$, and remain with a subgraph $T'$ that has fewer than $5N/2$ edges.
Next, we define a weight function $\mu$ over the edges of $T'$, such that every edge in $F$ has a weight of 1 and every other edge has
a weight of $1/2$. We define the weight $\mu(M_2)$ of a perfect matching $M_2 \subset T'$ as the product of the weights of its edges.
Therefore, if $M_2$ contains exactly $j$ edges of $F$, then $\mu(M_2) = (1/2)^{N/2-j}$. Moreover, for such a matching $M_2$, we have $|F\setminus M_2|=N/2-3-j$.
Clearly, $F\setminus M_2$ is also a set of ps-flippable edges, none of which belongs to $M_1 \cup M_2$. We thus have
\[\frac{1}{\supp{M_1 \cup M_2}} \le \frac{1}{2^{N/2-3-j}} = 8 \mu(M_2), \]
which implies that, given a specific triangulation $T$ and a specific perfect matching $M_1 \subset T$,
\begin{equation} \label{eq:pmVSsupport}
\sum_{M_2 \in \pmt{T'}} \frac{1}{\supp{M_1 \cup M_2}} \le 8 \sum_{M_2 \in \pmt{T'}} \mu(M_2),
\end{equation}
with $T'=T\setminus M_1$, as above.
\paragraph{Kasteleyn's technique: An enhanced version.} We now apply an extension of Kasteleyn's technique to estimate the sum in the right-hand side of (\ref{eq:pmVSsupport}).
Here is a brief overview of the technique being used (where instead of the original technique, we apply a weighted extension of it).
Given an oriented graph\footnote{We follow here the notation used in \cite{LP86} to denote a digraph obtained from an underlying undirected graph by giving each of its edges an orientation.} $\vec{G}=(V,E)$ with no anti-parallel edges and a weight function $\mu$ over the edges,
we define the following weighted adjacency matrix $B_{\vec{G},\mu} = (b_{ij})_{N\times N}$
of $(\vec{G},\mu)$,
\begin{equation*}
b_{ij} = \left\{
\begin{array}{ll}
\mu(e), & \text{if } e=(i,j)\in E,\\
-\mu(e), & \text{if } e=(j,i)\in E,\\
0, & \text{otherwise}
\end{array} \right.
\end{equation*}
(where $N=|V|$, and the rows and columns of $B_{\vec{G},\mu}$ correspond to an arbitrary fixed enumeration of the vertices).

An easy extension of Kasteleyn's theorem states that every planar graph $G$ can be oriented into some digraph $\vec{G}$ such that,
for any real-valued weight function $\mu$ on its edges, we have
\begin{equation} \label{eq:kast}
 \left( \sum_{M \in \pmt{G}} \mu(M) \right)^2 = \left| \mathsf{det}\left(B_{\vec{G},\mu}\right) \right|
\end{equation}
(recall that $\mu(M)=\prod_{e \in M}\mu(e)$).
In the ``pure" form of Kasteleyn's theorem $\mu \equiv 1$ (i.e., $G$ is unweighted) and the left-hand side is just the
squared number of perfect matchings in $G$. A detailed presentation of Kasteleyn's theorem can be found in \cite[Section 8.3]{LP86}.
The extension to weighted graphs is given in Exercise 8.3.9 therein.

We denote by $b_i$ the column vectors of $B$, for $1 \le i \le N$,
and estimate the above determinant using Hadamard's inequality
\begin{equation} \label{eq:had}
\left| \mathsf{det}\left(B_{\vec{G},\mu}\right) \right| \le \prod_{i=1}^{N}\|b_i\|_2.
\end{equation}
Applying the above machinery to our plane graph $T'$ (i.e., using \eqref{eq:kast} and \eqref{eq:had}), with the edge weights $\mu$ as defined above, we have
\begin{eqnarray}
\sum_{M_2 \in \pmt{T'}} \mu(M_2) & = & \sqrt{\left| \mathsf{det}(B_{\vec{T}',\mu}) \right|}  \le \sqrt{\prod_{i=1}^{N}\|b_i\|_2} = \left(\prod_{i=1}^{N}\|b_i\|_2^2\right)^{1/4} \nonumber \\[0.3em]
& \le & \left(\frac{1}{N}\sum_{i=1}^{N}\|b_i\|_2^2\right)^{N/4} = \left(\frac{2}{N} \sum_{e \in T'} \mu(e)^2 \right)^{N/4} \label{eq:kastel}
\end{eqnarray}
(where we have used the arithmetic-geometric mean inequality and the fact that every edge of $\vec{T}'$ has two corresponding matrix entries).
We note that the bound $6^{N/4}$ on the number of perfect matchings in a triangulation $T$ is obtained in \cite{BKKSS07} by applying the unweighted version of
Kasteleyn's theorem to the entire $T$. In this case $\sum_{e \in T}\mu(e)^2$ is the number of edges of $T$, which is at most $3N$, and the bound follows.

By noting that
\[|T'\setminus F| \le 5N/2 - (N/2-3) = 2N + 3, \]
 and combining this with (\ref{eq:pmVSsupport}) and (\ref{eq:kastel}), we obtain
\begin{eqnarray}
\sum_{M_2 \in \pmt{T'}} \frac{1}{\supp{M_1 \cup M_2}} & \le & 8\cdot \left(\frac{2}{N} \sum_{e \in T'} \mu(e)^2 \right)^{N/4} \nonumber \\[0.2em]
& \le & 8\cdot \left(\frac{2}{N} \cdot \left(1^2\cdot (N/2-3) + \left(\frac{1}{2}\right)^2 \cdot (2N + 3) \right) \right)^{N/4} \nonumber \\[0.2em]
& = & O\left(2^{N/4}\right). \label{eq:M2}
\end{eqnarray}
Recalling once again that a triangulation contains at most $6^{N/4}$ perfect matchings \cite{BKKSS07} (that is, there are $6^{N/4}$ ways of choosing $M_1$), and combining this with (\ref{eq:M2}), we obtain
\[ \scyc{S} \le \sum_{T \in \tr{S}} \sum_{M_1,M_2 \in \pmt{T} \atop M_1,M_2 \text{ edge-disjoint}} \frac{1}{\supp{M_1 \cup M_2}} \le \sum_{T \in \tr{S}} 6^{N/4} \cdot O\left(2^{N/4}\right)
= O\left(12^{N/4}\right) \cdot \tri{S}, \]
as asserted.
\end{theProof} \vspace{2mm}

Combining Theorem \ref{th:scVtr} with the bound $\tri{N} < 30^N$ \cite{SS10}, we obtain
\begin{corollary}
$\scyc{N} = O\left(55.8363^N \right)$.
\end{corollary}

\section{Perfect matchings and triangulations} \label{sec:matchings}
In this section we apply the machinery of the previous section to derive an upper bound on the ratio between the number of plane perfect matchings
and the number of triangulations. As already mentioned in Section \ref{sec:pre}, Kasteleyn's technique implies that a triangulation of a set of $N$ points
can contain at most $6^{N/4}$ perfect matchings (see \cite{BKKSS07}).
This implies that for every set $S$ of $N$ points in the plane, $\pmtc{S} \le 6^{N/4} \cdot \tri{S} \approx 1.5651^N \cdot \tri{S}$.
We will improve this bound, using lower bounds on the supports of perfect matchings, in a manner similar to that in Section \ref{se:firstB}.

Before proceeding, we note the following lower bound on the ratio $\pmtc{S} / \tri{S}$. Let $S$ be a \emph{double circle configuration},
depicted in Figure \ref{fi:triple}(b), consisting of $N$ points (see \cite{HuNo97} for a precise definition).
An inclusion-exclusion argument implies that $\tri{S} = 12^{N/2}$ (see \cite{HuNo97,SaSe03}).
Moreover, Aichholzer~{\em et al.}~\cite{AHHHKV07} proved that $\pmtc{S} = \Theta^*(2.2^N)$.
Therefore, in this case, $\pmtc{S} / \tri{S} \approx \Theta^*(0.635^N)$.

We now present an improved upper bound for this ratio.
\begin{theorem}
For any set $S$ of $N$ points in the plane,
\[\pmtc{S} \le 8 \cdot (3/2)^{N/4} \cdot \tri{S} = O(1.1067^N) \cdot \tri{S}. \]
\end{theorem}
\begin{theProof}{\!\!}
The exact value of $\pmtc{S}$ is
\begin{equation} \label{eq:PMsupp}
\pmtc{S} = \sum_{T \in \tr{S}} \sum_{M \in \pmt{T}}\frac{1}{\supp{M}}.
\end{equation}
Consider a triangulation $T \in \tr{S}$ and a perfect matching $M \subseteq T$. As shown in the proof of Theorem \ref{th:scVtr},
there exists a set of $N/2-3$ ps-flippable edges in $T \setminus M$.
Therefore, the support of $M$ is at least $2^{N/2-3}$. Combining this with (\ref{eq:PMsupp}) implies
\[\pmtc{S} \le \sum_{T \in \tr{S}} \sum_{M \in \pmt{T}}\frac{1}{2^{N/2-3}} \le \sum_{T \in \tr{S}} \frac{6^{N/4}}{2^{N/2-3}} = 8\cdot (3/2)^{N/4} \cdot \tri{S}. \]
\end{theProof}

As already mentioned above, this does not imply a new bound on $\pmtc{N}$, since  Sharir and Welzl \cite{SW06} showed that
$\pmtc{S} = O\left(10.05^N\right)$, bypassing the approach of counting matchings \emph{within} a triangulation.
We are not aware of any construction for which $\pmtc{S} \ge \tri{S}$, and offer the conjecture that there exists
a constant $c<1$ such that $\pmtc{S}=O(c^{|S|} \cdot \tri{S})$ for every finite set $S$ of points in the plane.
(See also the conjecture concerning spanning cycles, made in the introduction.)

\section{An improved bound} \label{sec:improved}
In this section we present a more complex analysis for the number of spanning cycles, obtaining a slightly better bound than the one presented in Section \ref{se:firstB}.
The analysis has three parts, each presented in a separate subsection.

Let us denote the number of interior vertices of degree 3 in the triangulation $T$ as $v_3(T)$. Moreover, let us denote the number of flippable
edges in $T$ as $\flip{T}$. In Subsection \ref{ssec:v3} we give an upper bound for $\sum_{C \in \scy{T}}\frac{1}{\supp{C}}$ that depends on $v_3(T)$.
In Subsection \ref{ssec:flip} we give an upper bound for $\sum_{C \in \scy{T}}\frac{1}{\supp{C}}$ that depends on $\flip{T}$.
Finally, in Subsection \ref{ssec:int} we combine these two bounds to obtain \[\scyc{N} = O(10.9247^{N/4}) \cdot \tri{N} = O(1.8181^N) \cdot \tri{N} =O(54.5430^N).\]

\subsection{A $v_3(T)$-sensitive bound} \label{ssec:v3}
In this subsection we derive the following bound, which is a function of $N$ and $v_3(T)$.
\begin{lemma} \label{le:v3upper}
Let $T$ be a triangulation over a set $S$ of $N \ge 6$ points in the plane, such that $N$ is even and $S$ has a triangular convex hull; also, let $v_3(T)=tN$. Then
\[\sum_{C \in \scy{T}}\frac{1}{\supp{C}} <  8\left(\frac{3}{2^t} \left( \frac{(2-t)(2-t/2)}{(1-t)^2} \right)^{1-t} \right)^{N/4}. \]
\end{lemma}
\begin{theProof}{\!\!}
As before, we treat every spanning cycle as the union of two edge-disjoint perfect matchings $M_1,M_2 \in \pmt{T}$.
We start by bounding the number of ways to choose the first perfect matching $M_1$.
For this, we use the standard variant of Kasteleyn's technique, with the weight function $\mu \equiv 1$ (i.e., $G$ is unweighted).

Recall the inequality $\sum_{M \in \pmt{T}} \mu(M) \le  \left(\prod_{i=1}^{N}\|b_i\|_2^2\right)^{1/4}$ obtained in Equation (\ref{eq:kastel}),
where the $b_i$ (for $1 \le i \le N$) are the column vectors of the adjacency matrix of the oriented graph $\vec{T}$.
Substituting $\mu \equiv 1$, the left hand side becomes the number of perfect matchings in $T$,  and the squared norm of each column vector is the degree of the vertex corresponding to that column.
Since every column that corresponds to a vertex of degree 3 has a squared norm of 3, the product of the squared norms of these is $3^{v_3(T)} = 3^{tN}$.

For the remaining $N-v_3(T)$ columns, we use, as in Section \ref{se:firstB}, the arithmetic-geometric mean inequality to bound the product of their squared norms (as in Equation (\ref{eq:kastel})).
This yields the bound
\begin{equation} \label{eq:NonV3}
\left(\frac{X}{N-v_3(T)}\right)^{(N-v_3(T))/4} = \left(\frac{X}{N(1-t)}\right)^{(N(1-t))/4},
\end{equation}
where $X$ is the sum of the degrees of all vertices other than those counted in $v_3(T)$. The sum of the degrees over the vertices of any specific triangulation is smaller than $6N$,
and the sum of the degrees of the interior degree-3 vertices in $T$ is $3v_3(T)$. Therefore, we have
\begin{equation} \label{eq:X}
X < 6N - 3v_3(T) = 3N(2-t).
\end{equation}
Combining (\ref{eq:NonV3}), (\ref{eq:X}), and the product of the squared norms that correspond to interior vertices of degree 3,
implies that the number of ways to choose $M_1$ is less than
\begin{equation} \label{eq:FirstPM2}
\left(3^t \cdot \left(\frac{3N(2-t)}{N(1-t)}\right)^{1-t} \right)^{N/4} = \left(3\cdot \left(\frac{2-t}{1-t}\right)^{1-t} \right)^{N/4}.
\end{equation}
Next, let us fix a specific perfect matching $M_1 \in \pmt{T}$.
As shown in the beginning of the proof of Theorem \ref{th:scVtr}, there exists a set $F$ of $N/2-3$ ps-flippable edges in $T$,
none of which belongs to $M_1$.

We continue as in the proof of Theorem \ref{th:scVtr}, by assigning a weight of 1 to the edges of $F$ and a weight of $1/2$ to the rest of the edges of $T\setminus M_1$,
and then applying Kaseteleyn's technique to bound the sum
\[ \sum_{M_2 \in \pmt{T'}} \frac{1}{\supp{M_1 \cup M_2}} \le 8 \sum_{M_2 \in \pmt{T'}} \mu(M_2) \le 8 \left( \prod_{i=1}^{N} \|b_i'\|_2^2\right)^{1/4},\]
where $b_i'$ are the column vectors of the oriented adjacency matrix of $T \setminus M_1$ (recall \eqref{eq:pmVSsupport}).

An interior vertex $v$ of degree 3 in $T$ has only two edges adjacent to it in $T\setminus M_1$, both not in $F$ (since an edge adjacent to an interior vertex of degree 3 cannot be flippable).
Therefore, the squared norm of a matrix column that corresponds to such a vertex is $(1/2)^2+(1/2)^2=1/2$,
and the product of the squared norms of all such columns is $1/2^{v_3(T)} = 1/2^{tN}$.

For the remaining $N-v_3(T)$ columns, we may once again use the arithmetic-geometric mean inequality to obtain a bound similar to the one in  \eqref{eq:NonV3}.
However, this time we get a different value for $X$, since (i) some of the edges of $T$ were removed, and (ii) some of the remaining edges were reweighted.
The edges of $F$ have remained and still have a weight of 1 each, so they contribute at most $2 \cdot (N/2-3) \cdot 1 < N$ to $X$.
Every other edge contributes $2 \cdot 1/4 = 1/2$ if it is not incident to an interior vertex of degree 3 in $T$, and $1/4$ otherwise.
Since a triangulation has fewer than $3N-3$ edges, there are fewer than $2N$ edges in $T\setminus\{F \cup M_1\}$, and we get
\[ X < N + (2N-2v_3(T)) \cdot \frac{1}{2} + 2v_3(T) \cdot \frac{1}{4} = 2N-\frac{v_3(T)}{2} = N(2-t/2). \]
By combining this with the rest of the squared norms and with the present version of  \eqref{eq:NonV3}, we have
\begin{equation} \label{eq:SecondPM}
\sum_{M_2 \in \pmt{T\setminus M_1}}\frac{1}{\supp{M_1 \cup M_2}} < 8 \left( \frac{1}{2^t} \cdot \left(\frac{N(2-t/2)}{N(1-t)}\right)^{1-t} \right)^{N/4}
\hspace{-2mm} = 8 \left( \frac{1}{2^t} \cdot \left(\frac{2-t/2}{1-t}\right)^{1-t} \right)^{N/4}.
\end{equation}
Finally, to complete the proof, we combine (\ref{eq:FirstPM2}) and (\ref{eq:SecondPM}), and obtain
\begin{eqnarray*}
 \sum_{C \in \scy{T}}\frac{1}{\supp{C}} \le \sum_{M_1,M_2 \in \pmt{T} \atop M_1,M_2 \text{ edge-disjoint}}\frac{1}{\supp{M_1 \cup M_2}} \quad\quad\quad\quad\quad\quad\quad\quad\quad \\[2mm]
 < \left(3\cdot \left(\frac{2-t}{1-t}\right)^{1-t} \right)^{N/4} \cdot 8 \left( \frac{1}{2^t} \cdot \left(\frac{2-t/2}{1-t}\right)^{1-t} \right)^{N/4}
 =  8 \left(\frac{3}{2^t} \left( \frac{(2-t)(2-t/2)}{(1-t)^2} \right)^{1-t} \right)^{N/4}.
\end{eqnarray*}
\end{theProof}

\paragraph{Remark.} Notice that in the worst case (i.e., when $t=0$) we obtain the same asymptotic value as in our initial bound of $12^{N/4}$.
Similarly, the bound in (\ref{eq:FirstPM2}) becomes $6^{N/4}$ when $t=0$, as in Buchin~{\em et al.}~\cite{BKKSS07}.

\subsection{A $\flip{T}$-sensitive bound} \label{ssec:flip}
Hurtado, Noy, and Urrutia \cite{HNU99} prove that that $\flip{T} \ge N/2-2$, and that this bound is tight in the worst case (the upper bound is also implied by Lemma \ref{le:psFlip}; see also \cite{HSSTW11}).
In this subsection we obtain a bound as a function of $\flip{T}$, which improves our initial bound of $12^{N/4}$ when $\flip{T}$ is
larger than $N/2$ by some positive fraction of $N$. We start by mentioning two basic properties of triangulated polygons.

\begin{figure}[h]
\centerline{\placefig{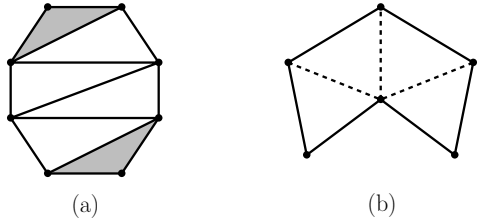}{0.38\textwidth}}
\vspace{-1mm}

\caption{\small \sf (a) The shaded faces are the two ears of the triangulated polygon.
(b) A polygon with three flippable diagonals, two of which form a ps-flippable set, and with $5=2^2(5/4)^1$ triangulations.}
\label{fi:part2}
\vspace{-2mm}
\end{figure}

\paragraph{Ears.} Given a triangulated polygon $P$, an \emph{ear} of $P$ is a bounded face of the triangulation with two of its edges on the boundary of $P$.
It can easily be shown that every triangulated simple polygon with at least four edges contains at least two ears (whose boundary edges are all distinct).
For example, the two ears of the triangulated polygon in Figure \ref{fi:part2}(a) are shown shaded.

\paragraph{Catalan numbers.} The $N$-th Catalan number is $\displaystyle C_N = \frac{1}{N+1} \binom{2N}{N}$.
It is well known that a convex polygon with $N$ vertices has $C_{N-2}$ triangulations (e.g., see \cite[Section 5.3]{St99}).
Therefore, the number of triangulations of a convex polygon with $d \ge 1$ diagonals is
\begin{equation} \label{eq:diag}
C_{d+1}= \frac{1}{d+2} \binom{2d+2}{d+1} \ge 2^d \left(\frac{5}{4} \right)^{d-1}.
\end{equation}
(the inequality can easily be verified by induction).
\vspace{5mm}

Next, we define $c_{\text{gon}}$ as the maximum real number satisfying the following property.
Every simple polygon $P$ that has a triangulation $T_P$ with $k$ of its diagonals flippable and with $l \le k$ of these diagonals forming a ps-flippable set,
has at least $2^l c_{\text{gon}}^{k-l}$ triangulations.
Notice that the triangulations under consideration, including $T_P$, are triangulations of the polygon $P$, and not of its vertex set.

\begin{lemma} \label{le:cgon}
$\displaystyle x \le c_{\text{gon}} \le \frac{5}{4}$ with $x \approx 1.17965$ the unique real root of the polynomial $1+4x^2-4x^3$.
That is, every simple polygon $P$ that has a triangulation $T_P$ with $k$ of its diagonals flippable and with $l \le k$ of these diagonals forming a ps-flippable set,
has at least $2^l x^{k-l}$ triangulations.
\end{lemma}
\begin{theProof}{\!\!}
Figure \ref{fi:part2}(b) depicts a polygon that implies the upper bound.

We prove the lower bound by induction on $l$ and $k$.
To have some base case for this induction, notice that when $l=k$, $P$ has at least $2^k = 2^l x^{k-l}$ triangulations.

Next, consider a polygon $P$ and a triangulation $T_P$ of $P$, such that $T_P$ contains a maximal set $F$ of $l$ ps-flippable edges, and $k-l > 0$ flippable edges not in $F$.
Let $f_1,f_2, \cdots, f_j$ be the non-triangular faces of the convex subdivision $T_P \setminus F$.
(Notice that $T_P \setminus F$ is a convex subdivision of a polygon, and not of the convex hull of a point set.)
Each $f_i$ is a convex polygon with $m_i \ge 4$ sides and $m_i-3$ diagonals, and thus, $l = \sum_{i=1}^j(m_i-3)$.

A flippable edge not in $F$ must be on the boundary of some $f_i$, since otherwise $F$ is not maximal.
We say that such a flippable edge $e \notin F$ \emph{is covered by a face} $f_i$ if (i) $e$ is on the boundary of $f_i$,
and (ii) the triangle $\Delta$ in $T_P$ that is contained in $f_i$ and incident to $e$ has two edges in $F$ (so $e$ is the only edge of $\Delta$ on the boundary of $f_i$).
Ears are incident to two edges of their containing $f_i$, and thus, cannot cover any edge. Since any triangulated polygon contains at least
two ears, with four distinct boundary edges, a polygon with $m$ sides (and $m-3$ diagonals) can cover at most $m-4$ edges.
Therefore, if all flippable edges not in $F$ are covered, then $k-l \le \sum_{i=1}^j(m_i-4)$.
By multiplying the number of triangulations of the $f_i$'s and applying (\ref{eq:diag}), we get that the number of triangulations of $P$ that contain $T_P \setminus F$ is at least
\[ \sum_{i=1}^{j} 2^{m_i-3}\left( \frac{5}{4} \right)^{m_i-4} = 2^{\sum_{i=1}^{j} (m_i-3)}\left( \frac{5}{4} \right)^{\sum_{i=1}^{j} (m_i-4)}
\ge 2^l \left( \frac{5}{4} \right)^{k-l} \ge 2^lx^{k-l}. \]

\begin{figure}[h]
\centerline{\placefig{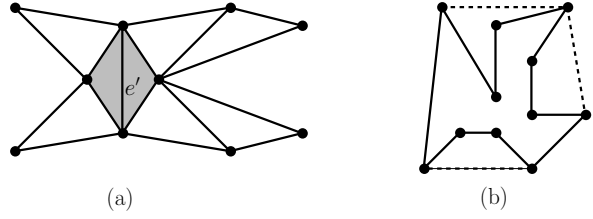}{0.47\textwidth}}
\vspace{-1mm}

\caption{\small \sf (a) Removing the two (shaded) triangles incident to $e'$ produces at most four triangulated sub-polygons.
(b) A spanning cycle partitions the convex hull of the point set into interior-disjoint polygons.}
\label{fi:part2b}
\vspace{-2mm}
\end{figure}

We are left with the case where there is a flippable edge $e \notin F$ that is not covered by any $f_i$.
Let $e'$ be the edge obtained by flipping $e$. We now derive a lower bound on the number of triangulations of $P$ that contain $e$ and on the number of triangulations of $P$ that contain $e'$.
To bound the number of triangulations that contain $e$, we partition $P$ into two interior disjoint simple polygons $P',P''$ by ``cutting" $P$ at $e$.
More precisely, we consider the two triangulated polygons $T_{P'},T_{P''} \subset T_P$.
Together, these two triangulated polygons contain $k-1$ diagonals that are flippable.
Moreover, the set $F$ remains a set of $l$ ps-flippable edges which are diagonals of the two polygons.
Thus, the induction hypothesis implies that there are at least $2^lx^{k-1-l}$ triangulations of $P$ that contain $e$.
To obtain a similar bound for the number of triangulations containing $e'$, we produce at most four triangulated sub-polygons of $T_P$ by removing
the two triangles incident to $e$ (which cover the same quadrilateral covered by the two triangles incident to $e'$ in the new triangulation). Such a case is illustrated in Figure \ref{fi:part2b}(a), where the triangles incident to $e'$ are shaded.
This partitioning may cancel the flippability of at most five edges (those incident to the two triangles adjacent to $e$).
At most two out of the five edges may be ps-flippable, since $e$ can only be incident to ears of polygons of $T_P \setminus F$.
Using the induction hypothesis again, we get that there are at least $2^{l-2}x^{(k-5)-(l-2)} = 2^{l-2}x^{k-l-3}$ triangulations that contain $e'$ (and the four edges around it).
Therefore, $P$ has at least
\[ 2^lx^{k-1-l} +  2^{l-2}x^{k-l-3} = 2^lx^{k-l} \left( \frac{1}{x} + \frac{1}{4x^3} \right) = 2^lx^{k-l} \cdot \frac{4x^2+1}{4x^3} = 2^lx^{k-l} \]
triangulations, as asserted (recall that $x$ is the root of the polynomial $1+4x^2-4x^3$).
\end{theProof} \vspace{2mm}

Next, we show how to use $c_{\text{gon}}$ and $\flip{T}$ to bound $\sum_{C \in \scy{T}}\frac{1}{\supp{C}}$.

\begin{lemma} \label{le:cgonUse}
Consider a triangulation $T$ with $\flip{T} = N/2 - 3 + \kappa N$, for some $\kappa \ge 0$, and let $x$ be the constant presented in Lemma \ref{le:cgon}. Then
\[ \sum_{C \in \scy{T}}\frac{1}{\supp{C}} < 8\left( \frac{(3+(\gamma^2-1)(\kappa+1/2))(4+(x^2-1)\kappa)}{x^{4\kappa}}  \right)^{N/4}, \]
where
\[ \gamma = x \cdot e^{-\frac{x^2-1}{4(4+(x^2-1)\kappa)}}. \]
\end{lemma}
\begin{theProof}{\!\!}
Once again, we treat every spanning cycle as the union of a pair of edge-disjoint perfect matchings $M_1,M_2 \in \pmt{T}$,
and use Kasteleyn's technique (as presented in Section \ref{se:firstB}) to bound the number of such pairs.
We start by fixing some perfect matching $M_1 \in \pmt{T}$ and denote the number of flippable edges of $T$ that are in $M_1$ as $\flipt{M_1}$.
As shown in the proof of Theorem \ref{th:scVtr}, there is a set of at least $N/2-3$ ps-flippable edges in $T\setminus M_1$.
We restrict our attention to a set $F$ of \emph{exactly} $N/2-3$ ps-flippable edges in $T\setminus M_1$.

For the choice of $M_2$, we define a weight function $\mu(\cdot)$ on the edges of $T\setminus M_1$, such that
\begin{equation*}
\mu(e) = \left\{
\begin{array}{ll}
2, & \text{if } e \in F,\\
x, & \text{if } e\notin F \text{ is flippable},\\
1, & \text{if } e \text{ is not flippable.}
\end{array} \right.
\end{equation*}
Notice that any spanning cycle partitions the convex hull of its point set into interior-disjoint simple polygons;
an example of such a partition is illustrated in Figure \ref{fi:part2}(b).
The support of the spanning cycle is the product of the number of triangulations of each of these polygons.
For a fixed choice of $M_2$ (and of $M_1$), denote by $P_1, \ldots, P_m$ the polygons in the partition produced by $M_1 \cup M_2$.
For each $i$, let $k_i$ be the number of flippable diagonals of $P_i$, and let $l_i$ be the number of those diagonals (among the $k_i$ flippable ones) that belong to $F$.
If $M_2$ uses $\flipt{M_2}$ flippable edges of $T\setminus M_1$, $l$ of which are in $F$, then $\sum_{i=1}^{m}k_i = \flip{T} - \flipt{M_1}-\flipt{M_2}$ and
 $\sum_{i=1}^{m}l_i =|F| -  l = N/2-3-l$. Applying Lemma \ref{le:cgon} to each $P_i$ and multiplying the resulting bounds, we obtain a total of at least
 $\displaystyle 2^{\sum l_i}x^{\sum k_i - \sum l_i}$ triangulations.
 Hence,
\[  \supp{M_1\cup M_2} \ge (2/x)^{\sum l_i}x^{\sum k_i} = (2/x)^{N/2-3-l} x^{\flip{T} - \flipt{M_1}-\flipt{M_2}}.   \]
Next, notice that $\mu(M_2) = 2^lx^{\flipt{M_2}-l}$, so we have
\begin{equation} \label{eq:flipSupp}
 \supp{M_1\cup M_2} \ge \frac{2^{N/2-3}x^{\flip{T}-\flipt{M_1}-(N/2-3)} }{\mu(M_2)} = \frac{2^{N/2-3}x^{\kappa N-\flipt{M_1}} }{ \mu(M_2)}.
\end{equation}
By combining (\ref{eq:flipSupp}) with Kasteleyn's method, we obtain
\begin{align}
\sum_{M_2 \in \pmt{T\setminus M_1}} \frac{1}{\supp{M_1 \cup M_2}} &\le \sum_{M_2 \in \pmt{T\setminus M_1}} \frac{\mu(M_2)}{2^{N/2-3}x^{\kappa N-\flipt{M_1}}} \nonumber \\[0.2em]
&\le \frac{1}{2^{N/2-3}x^{\kappa N-\flipt{M_1}}} \cdot \left( \frac{2}{N} \sum_{e\in T\setminus M_1 }\mu(e)^2 \right)^{N/4}. \label{eq:M2flip}
\end{align}
To bound the sum in the parentheses, we write $\flipt{M_1} = w+z$, where $w=|F \cap M_1|$ and $z$ is the number of flippable edges in $M_1 \setminus F$. Then $T\setminus M_1$ contains fewer than $N/2 - w$ edges of $F$, $\kappa N - z$ flippable edges not in $F$, and fewer than $2N-(\kappa N - \flipt{M_1})$ non-flippable edges.
Therefore,
\[\frac{2}{N} \sum_{e\in T\setminus M_1 }\mu(e)^2 <  \frac{2}{N} \left( 1^2 \cdot \left(2N-(\kappa N - \flipt{M_1})\right) + x^2 \cdot (\kappa N - z)+ 2^2 \cdot (N/2-w)  \right). \]
Since $x<2$, the right-hand side is maximized when $w=0$ and $z=\flipt{M_1}$, and it then becomes
\begin{eqnarray}
\frac{2}{N} \sum_{e\in T\setminus M_1 }\mu(e)^2 & < & \frac{2}{N} \left( 1^2 \cdot \left(2N-(\kappa N - \flipt{M_1})\right) + x^2 \cdot (\kappa N - \flipt{M_1})+ 2^2 \cdot N/2  \right) \nonumber \\[0em]
& = & 8 + 2(x^2-1)\kappa - 2(x^2-1)\cdot \flipt{M_1}/N \nonumber \\[0.3em]
& = & (8+2(x^2-1)\kappa) \left(1 - \frac{2(x^2-1)}{8+2(x^2-1)\kappa} \cdot \frac{\flipt{M_1}}{N}   \right) \nonumber \\[0.2em]
& \le & (8+2(x^2-1)\kappa) \cdot e^{-\frac{x^2-1}{4+(x^2-1)\kappa}\cdot \frac{\flipt{M_1}}{N}} \quad \quad \quad \qquad  \text{(using $1-u \le e^{-u}$ for $u\ge 0$)} \nonumber \\[0.2em]
& = & (8+2(x^2-1)\kappa) \cdot \left(\gamma / x \right)^{4 \cdot \flipt{M_1}/N}.   \label{eq:edgesum}
\end{eqnarray}
Combining (\ref{eq:M2flip}) and (\ref{eq:edgesum}), we get
\begin{align}
\sum_{C \in \scy{T}}\frac{1}{\supp{C}} &\le \sum_{M_1,M_2 \in \pmt{T} \atop M_1,M_2 \text{ edge-disjoint}} \frac{1}{\supp{M_1 \cup M_2}} \nonumber \\[0.2em]
&\le  \sum_{M_1 \in \pmt{T}} \frac{1}{2^{N/2-3}x^{\kappa N-\flipt{M_1}}} \left( (8+2(x^2-1)\kappa) \cdot \left(\gamma / x \right)^{4 \cdot \flipt{M_1}/N} \right)^{N/4} \nonumber \\[0.3em]
&= 8 \left( \frac{8+2(x^2-1)\kappa}{4x^{4\kappa}} \right)^{N/4} \sum_{M_1 \in \pmt{T}}\gamma^{\flipt{M_1}}. \label{eq:almostSCflip}
\end{align}
To bound the sum in (\ref{eq:almostSCflip}), we once again use Kasteleyn's technique.
This time, we define a weight function $\nu(\cdot)$ over the edges of $T$, such that every flippable edge gets a weight of $\gamma$, and every other edge a weight of 1.
Notice that, in this manner, $\nu(M_1)=\gamma^{\flipt{M_1}}$ for every $M_1 \in \pmt{T}$. We thus have
\begin{eqnarray}
\sum_{M_1 \in \pmt{T}}\gamma^{\flipt{M_1}} & \le & \left( \frac{2}{N} \sum_{e\in T} \nu(e)^2 \right)^{N/4} \nonumber \\[0.2em]
& < & \left( \frac{2}{N} \left(\gamma^2\cdot \flip{T} + 1 \cdot (3N-\flip{T})\right) \right)^{N/4} \nonumber \\[0.2em]
& < & \left( 6+ \frac{2}{N} (\gamma^2-1)(N/2 + \kappa N)  \right)^{N/4} \nonumber \\[0.2em]
& < & \left( 6 + 2(\gamma^2-1)(\kappa + 1/2) \right)^{N/4}. \label{eq:M1flip}
\end{eqnarray}
Finally, combining (\ref{eq:almostSCflip}) and (\ref{eq:M1flip}) implies the assertion of the lemma.
\end{theProof} \vspace{2mm}

Note that in the worst case, when $\kappa=0$, the bound becomes $O\left( (10+2\gamma^2)^{N/4} \right)$. For $k=0$, we have $\gamma = x \cdot e^{-(x^2-1)/16}$,
and it is easy to verify that $\gamma > 1$ for $1< x \le 5/4$. So the bound is actually asymptotically \emph{worse} than our initial bound of $12^{N/4}$,
and it continues to be worse when $\kappa$ is sufficiently small.
As the next subsection shows, in this case the $v_3$-dependent bound from Subsection \ref{ssec:v3} becomes small and can be used instead.

\subsection{Integration} \label{ssec:int}
In this subsection we combine the results from the two previous subsections to obtain an improved bound for $\scyc{N}$.
This is done by deriving a connection between $v_3(T)$ and $\flip{T}$. We start by presenting a generalization of Lemma \ref{le:even}.

\begin{lemma} \label{le:evenTri}
Let $c>1$ be a constant such that every set $S$ of an even number of points in the plane and a triangular convex hull satisfies $\scyc{S} = O(c^{|S|})$.
Then $\scyc{S} = O(c^{|S|})$ also holds for every other finite point-set $S$ in the plane.
\end{lemma}
\begin{theProof}{\!\!}
Consider a point set $S$. If $S$ has an even number of points, we pick a new point $p$ outside the convex hull of $S$, and denote $S' = S \cup \{p\}$.
As mentioned in the proof of Lemma \ref{le:even}, inserting an additional vertex outside the convex hull of the point set can only increase the number of spanning cycles.
If $S$ has an odd number of points, we denote $S'=S$.
Notice that, either way, $S'$ has an odd number of points.
Let $\Delta abc$ be a large triangle containing $S'$ in its interior, and let $S''=S' \cup \{a,b,c\}$.
Again, since inserting an additional vertex outside the convex hull of the point set can only increase the number of spanning cycles, we have $\scyc{S'} \le \scyc{S'\cup \{a\}} \le \scyc{S'\cup \{a,b\}} \le \scyc{S''}$. Since $S''$ has an even number of points and
a triangular convex hull, $\scyc{S} \le \scyc{S'} \le \scyc{S''} = O(c^{N+3}) = O(c^{|S|})$.
\end{theProof}

\begin{figure}[h]
\centerline{\placefig{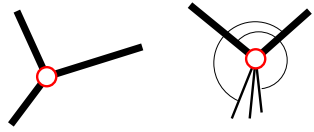}{0.25\textwidth}}
\vspace{-1mm}

\caption{\small \sf Separable edges.}
\label{fi:SepEdges}
\vspace{-2mm}
\end{figure}

We also require the notion of \emph{separable edges}, as presented in \cite{SSW10}.
Consider a point set $S$, a triangulation $T \in \tr{S}$, and an interior point $p \in S$.
We call an edge $e$ incident
to $p$ in $T$ a  {\em separable edge at $p$} if it can be
separated from the other edges incident to $p$ by a line
through $p$.  An equivalent condition is
that the two angles between $e$ and its clockwise and
counterclockwise neighboring edges (around $p$) sum up to
more than $\pi$.
We observe the easy following properties
(see Figure \ref{fi:SepEdges} for an illustration).
\begin{itemize}
\item[(S0)]
No edge is separable at both vertices induced by its endpoints.
\item[(S1)]
If $p$ has degree 3 in $T$, every edge incident to it is separable
(recall that $p$ is an interior point).
\item[(S2)]
If $p$ has degree at least $4$ in $T$, at most two incident edges can
be separable at $w$.
\item[(S3)]
If $p$ has degree at least $4$ in $T$ and there are two edges
separable at $p$, then they must be consecutive in the order around it.
\end{itemize}

We are now ready for our main theorem.

\begin{theorem} \label{th:impSC}
For any set $S$ of $N$ points in the plane,
\[ \scyc{S} = O\left(10.9247^{N/4}\right) \cdot \tri{S} = O\left(1.8181^N\right) \cdot \tri{S}. \]
\end{theorem}
\begin{theProof}{\!\!}
By Lemma \ref{le:evenTri}, we may assume that $N$ is even and that $S$ has a triangular convex hull.
Recall that
\[ \scyc{S} = \sum_{T \in \tr{S}} \sum_{C \in \scy{T}}\frac{1}{\supp{C}}. \]
We sort the triangulations in the first sum according to the number of interior vertices of degree 3 that they contain, and get
\begin{equation} \label{eq:scByi}
\scyc{S} = \sum_{i=0}^{(2N+1)/3}\sum_{T \in \tr{S} \atop v_3(T)=i} \sum_{C \in \scy{T}}\frac{1}{\supp{C}}.
\end{equation}
(The fact that $v_3(T) \le (2N+1)/3$ for every triangulation $T$ is established, e.g., in \cite{SW06b}.)
Given a triangulation $T$ with $v_3(T)=i$, we can use Lemma \ref{le:v3upper} to bound $\sum_{C \in \scy{T}}\frac{1}{\supp{C}}$.
However, when $v_3(T)$ is small, the improvement in Lemma \ref{le:v3upper} is not significant.
In this case we will use instead the bound in Lemma \ref{le:cgonUse} which, as we now proceed to show, becomes significant when $v_3(T)$ is small.

Consider a triangulation $T \in \tr{S}$. Since $S$ has a triangular convex hull, $T$ contains $3N-9$ interior edges.
Notice that an interior edge $e$ is flippable if and only if $e$ is not separable at either of its endpoints (this property is equivalent to
$e$ being a diagonal of a convex quadrilateral). From the above properties of separable edges, we have
\[ \flip{T} \ge \overbrace{3N-9}^{\text{Interior edges}} - 3\cdot\overbrace{v_3(T)}^{\text{Interior vertices of degree 3}} - 2\cdot\overbrace{(N-v_3(T)-3)}^{\text{Other interior vertices}} =
N - 3 - v_3(T). \]
To find for which values of $i$ it is better to use Lemma \ref{le:v3upper}, and for which values it is better to use Lemma \ref{le:cgonUse},
we define $t = v_3(T) / N$ and
\[ \kappa=\frac{\flip{T}-(N/2-3)}{N} \ge \frac{(N-tN-3) - (N/2 -3) }{N} = 1/2-t, \]
and solve the equation
\[ 8\left(\frac{3}{2^t} \left( \frac{(2-t)(2-t/2)}{(1-t)^2} \right)^{1-t} \right)^{N/4} = 8\left( \frac{(3+(\gamma^2-1)(\kappa+1/2))(4+(x^2-1)\kappa)}{x^{4\kappa}}  \right)^{N/4}, \]
where $x \approx 1.17965$ and $\gamma = x \cdot e^{-\frac{x^2-1}{4(4+(x^2-1)\kappa)}}$; this will determine the threshold where the two bounds coincide.
That is, we need to solve the equation
\[ \frac{3}{2^t} \left( \frac{(2-t)(2-t/2)}{(1-t)^2} \right)^{1-t} = \frac{(3+(\gamma^2-1)(\kappa+1/2))(4+(x^2-1)\kappa)}{x^{4\kappa}}. \]
For this, we use the Wolfram Mathematica software \cite{Wolfram09}, and obtain the solution $t \approx 0.1072$.
Moreover, it is easily shown that for $i \ge 0.1072N$ the bound from Lemma \ref{le:v3upper} is smaller, and for $i \le 0.1072N$ the bound from Lemma \ref{le:cgonUse} is smaller.
In fact, these bounds, in their appropriate usage, are all dominated by the common bound for $t \approx 0.1072$. This, together with (\ref{eq:scByi}), imply the asserted bound.
\end{theProof}

By combining Theorem \ref{th:impSC} with the bound $\tri{N}<30^N$ \cite{SS10}, we obtain:
\begin{corollary}
$\scyc{N} = O\left(54.5430^N \right)$.
\end{corollary}



\end{document}